\documentclass{article}
\usepackage[english]{babel}
\usepackage[margin=1in]{geometry}
\usepackage{multicol}
\usepackage{background}
\usepackage{graphicx}
\usepackage{float}
\usepackage{multicol,caption}
\usepackage{cite}

\usepackage{amsmath,array,graphicx}
\usepackage{verbatim}
\usepackage[font=small,labelfont=bf]{caption}
\newenvironment{Table}
   {\par\bigskip\noindent\minipage{\columnwidth}\centering}
   {\endminipage\par\bigskip}

\newenvironment{Figure}
  {\par\medskip\noindent\minipage{\linewidth}}
  {\endminipage\par\medskip}
  
\usepackage[math]{blindtext}

\SetBgScale{4}
\SetBgColor{gray}
\SetBgAngle{90}
\SetBgContents{arXiv: some other text goes here}
\SetBgPosition{-1.5,-10}

\begin{document}
 
{\centering
 
{\bfseries\Large Development and Experimental Validation of a Viscosity Meter for Newtonian and Non-Newtonian Fluids \bigskip}
 
Ra\'ul O. Rojas \textsuperscript{1}, Juan C. Quijano\textsuperscript{1}, Claudia P. Tavera Ruiz\textsuperscript{2} and Alex F. Estupi\~n\'an L.\textsuperscript{2} \\ 
   
   {\vspace{0.4cm} \itshape

\textsuperscript{1}Universidad Aut\'onoma de Bucaramanga (UNAB), Programa de Ingenier\'ia Biom\'edica, Departamento de Matem\'aticas y Ciencias Naturales, Santander, Colombia. \\
\textsuperscript{2}Universidad de Investigaci\'on y Desarrollo (UDI), Departamento de Ciencias B\'asicas y Humanas, Santander, Colombia. \\

\vspace{0.2cm}

\normalfont (Dated: August 14, 2020)
 
   }
}
 
\begin{abstract}
\noindent The study of viscosity, in the area of fluid physics at a university level, is of great importance because of the various applications that are presented in the different fields of engineering. In this work an experimental method of implementation and validation is exposed, to be able to calculate the viscosity of some newtonian and non-newtonian fluids, in which the method of a sphere that descends through a fluid has been used, we implemented a viscometer of our own construction, with the help of the CassyLab sensor and software of Leybold Didactics, we show the results obtained by our measuring instrument, which is intended to highlight the versatility and precision of the measuring instrument prepared by us, in addition. In this research the authors want to motivate the physics laboratory teachers; to explore the use of these tools that allow you to check the topics seen in the theoretical classes. Finally, we present the hardworking results of the measurement of viscosity for different fluids, both newtonian and non-newtonian, for the latter we show the viscosity behavior as a function of temperature.
\bigskip
 
 
\end{abstract}
 
\begin{multicols}{2}

\section{Introduction}

One of the great applications of the calculation of the viscosity of the fluids, is directed mainly to the area of electrochemical research, in which it is sought to relate the electrical conductivity with the viscosity of the fluid to be studied, one of the most studied fluids in this Last decade are vegetable oils \cite{ref_1,ref_11} and milk \cite{ref_2} for different temperatures. Among the most interesting findings, they showed that proteins and lactose affected the electrical conductivity of milk by modifying its viscosity, in most liquid foods the concentration of electrolytes is relatively low, with soy sauce and fish sauce \cite{ref_3}.

Another field in which the viscosity of the fluids has a great application is civil engineering, which consists in determining the fluidity state of the asphalts at the temperatures used during their application \cite{ref_31,ref_32}, for example: The viscosity is measured in the Saybolt-Furol viscosity test or in the kinematic viscosity test. 

The viscosity of an asphalt cement at the temperatures used in mixing (normally 135 C) is measured with capillary flow meters viscometers or Saybolt viscometers; The absolute viscosity, at high operating temperatures (60 C), is usually measured with vacuum glass capillary viscometers \cite{ref_4, ref_5}. 

We, too, can see the importance of viscosity in tribology, such as that science that is responsible for studying friction, which generates wear and lubrication in some materials, as occurs in some mechanical mechanisms such as teeth of a gear. 

Where it should be taken into account that at higher contact velocitys of these materials, a hydrodynamic (or elastohydrodynamic) lubrication film is formed in highly charged mechanisms where the surfaces in contact are deformed by the action of these forces.

At this stage the minimum wear values are achieved because the oil or grease that was designed for this friction regime acts. Showing that the higher the viscosity of the lubricant, we will have greater losses due to viscous friction, so lubricants with high viscosity oils are only recommended in slow-moving elements \cite{ref_6, ref_7}.

In this work, we present the results obtained from the experimental implementation, which was carried out for the measurement of the dynamic viscosity of Newtonian fluids such as sunflower oil, glycerine and non-Newtonian fluids such as yogurt, cornstarch and ketchup, for each of the above fluids varying the temperature. 

In addition, we show the analytical development, using two different methods, to show the validity of our experiment and the quality of the data taken in it.

\section{Theoretical framework}

The classic definition for viscosity, arises as internal friction force that brought in a fluid. Where it should be taken into account, that these viscous forces oppose the movement of a portion of the fluid in relation to another. Fluids that flow easily, such as water and gasoline, have lower viscosity than thick liquids such as honey or oil. The viscosity of all fluids are very dependent on temperature, increase for gases and decrease for liquids as the temperature rises. A viscous fluid tends to adhere to a solid surface that is in contact with it.

In order to better understand the definition of viscosity of a fluid, we place a rectangular section, on some surface free of an unconfined fluid. We hope that on this surface a shear stress is experienced or appears in the direction parallel to the surface of the fluid. In addition, a gradient of velocities will appear in the fluid as a result of said shear stress, the velocity of the sheet being equal to that of the particles in contact with it (adhesion condition), as can be seen in Figure \ref{fig_1}.

\begin{Figure}
 \centering
 \includegraphics[width=0.5\linewidth]{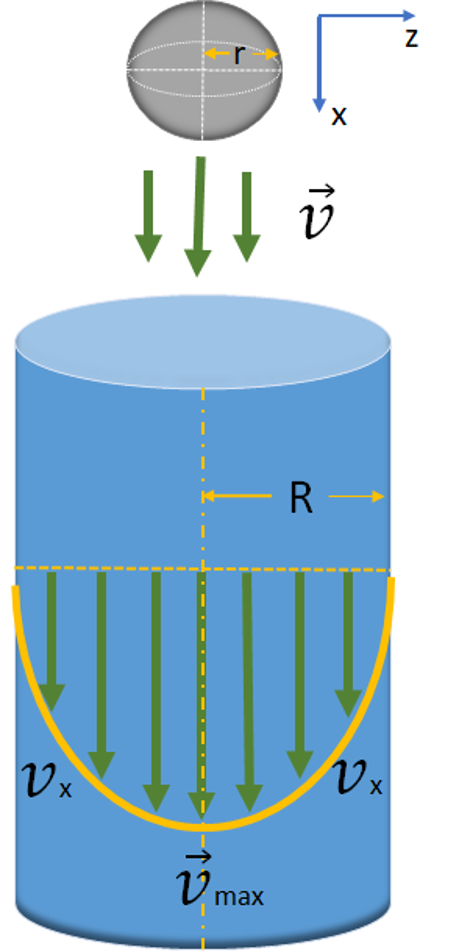}
 \captionof{figure}{Schematic illustration of the velocity profile defined from Newton's vision.}
    \label{fig_1}
\end{Figure}

There is one method to determine the viscosity of a fluid, using a falling sphere through the fluid. The method consists in release the sphere near the top surface of fluid, in order to ensure free falling and constant velocity during the journey. 

While the sphere is falling, it´s experiencing the action of three forces: its weight or gravitational force $F_g$, the buoyant force $F_b$ and the viscosity force $F_v$. If the movement is considered with constant velocity, the sum of these forces is equal to zero.The buoyant force depends on: fluid density $\rho_f$, gravitational acceleration $g$, and the volume of the sphere $V$ (See Figure \ref{graf_2}). Besides, the force due the viscosity depends on radius $r$ and velocity $v$ of the sphere, and the fluid´s viscosity $\eta$. Moreover, $F_g$ can be written in terms of the object density $\rho_{obj}$ and volume $V$, as seen below (See Equation (\ref{eq_1})) \cite{ref_fig_1,libro_fluidos}:     

\vspace{-0.5cm}

\begin{eqnarray}
    \nonumber F_b+F_v=F_g \\
    \nonumber m_s g + F_v=mg \\
    \nonumber \rho_f g V + 6\pi\eta rv=\rho_{obj} gV\\
    \nonumber 6\pi\eta rv=(\rho_{obj}-\rho_f)gV \\
    6\pi\eta rv= \Delta \rho g \frac{4}{3}\pi r^3
    \label{eq_1}
   \end{eqnarray}
 Solving for $\eta$ we have the Equation (\ref{eq_2}):
\begin{eqnarray}   
    \eta=\frac{2\Delta \rho g r^2}{9v}
    \label{eq_2}
\end{eqnarray}

\begin{Figure}
 \centering
 \includegraphics[width=\linewidth]{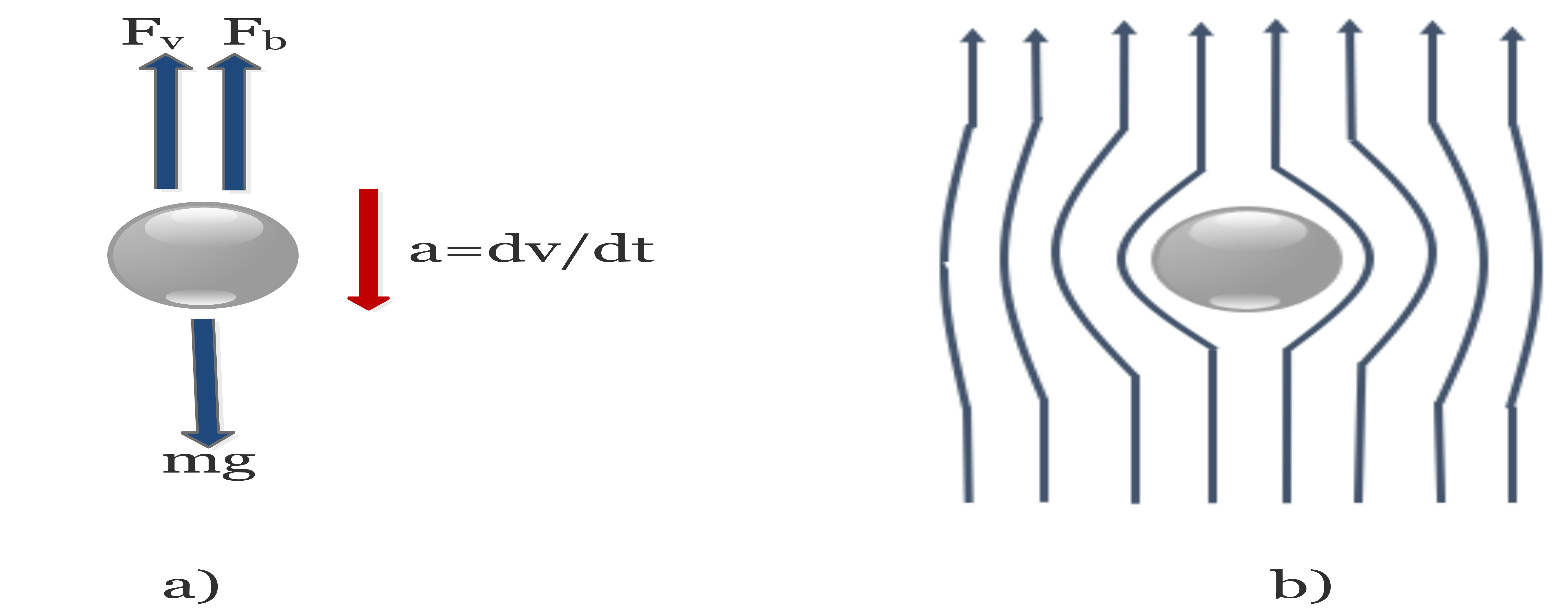}
 \captionof{figure}{a.) Free body diagram for the sphere inside the fluid. b.) Schematic representation of the laminar flow disturbance present in the fluid, due to the presence of the sphere.}
    \label{graf_2}
\end{Figure}

The previous formula applies in case of an infinitely extended fluid, but according to the experiment a correction factor is need to be added (See Equation (\ref{eq_3})) \cite{ref_libro,libro_2}. 

\begin{equation}
        \eta=\frac{2\Delta \rho g r^2}{9v} \left( \frac{1}{1+2.4 r/R} \right)
        \label{eq_3}
\end{equation}

Where $R$ is the radius of the test tube. Stokes' law is subject to a restriction in terms of its use and involves considering a laminar flow. Laminar flow is defined as that condition, in which fluid particles move along the smooth paths in the sheet, in other words, is when an orderly and smooth movement of the particles that form the fluid occurs.

To predict the type of flow that a fluid will present in a cross-section pipe, the Reynolds number should be calculated. This dimensionless number is a ratio between inertial and frictional force, and takes different expressions depending on whether it is for a pipe with a non-circular transversal section, open channels or fluid flows around a body \cite{cl_1}. In our experiment, it is a sphere submerged in a fluid moving with velocity V, in this case the Reynolds number can be calculated experimentally by the Equation (\ref{reinolds}) \cite{cl_2}.

\begin{equation}
    Re = \frac{\rho_f v_s r}{\mu}
    \label{reinolds}
\end{equation}

Where $\rho_f$ is the fluid density, $V_s$ the sphere velocity, r the sphere radius and µ the viscosity calculated experimentally. If the calculated $Re<1$ then the fluid will present a laminar flow and, if $Re>1$ the flow is turbulent.

In order to know the viscous behavior of a fluid, it is necessary to determine the shear stress and the velocity gradient. The main equations to determine the shear stress are the following \cite{cl_3, cl_4}: 

\begin{equation}
  \mathit{Newton's} \hspace{0.1cm} \mathit{law:} \quad
  \tau = \mu \left( \frac{dv}{dz} \right)   
  \label{eq_newton}
\end{equation}

\begin{equation}
  \mathit{Power} \hspace{0.1cm} \mathit{law:} \quad
  \tau = k \left( \frac{dv}{dz} \right)^{n}   
  \label{power's_law}
\end{equation}

\begin{equation}
  \mathit{Bingham's} \hspace{0.1cm} \mathit{equation:} \quad
  \tau = \tau_0 + \eta'  \left( \frac{dv}{dz} \right)
\end{equation}

\begin{equation}
  \mathit{Herschel-Bulkley's} \hspace{0.1cm} \mathit{ model:} \quad
  \tau = \tau_0 + k_H  \left( \frac{dv}{dz} \right)
\end{equation}

Where $\tau$ is the shear stress, $\eta$ viscosity, $\left( \frac{dv}{dz} \right)$  the velocity gradient, $k$ the consistency index, $n$ is the flow behavior index, $\tau_0$ the creep threshold, $\eta'$   plastic viscosity and $k_H$ consistency index for Herschel-Bulkley's fluids.

Depending on the effect of shear stress on the fluid, these can be classified as Newtonian and non-Newtonian fluids. Therefore, the mathematical model to be used to determine the shear stress depends on the type of fluid to be used. 

Newtonian fluids are characterized because their rheological behavior can be described by Newton's law (Equation (\ref{eq_newton})). This means that the shear stresses required to achieve a velocity are always linearly proportional, having a constant viscosity \cite{cl_6}. 

On the other hand, non-Newtonian fluids can not be described by Newton's law. In this case, viscosity is no longer talked about, because the ratio between shear stress and velocity is not constant. That viscosity function as a function of velocity is known as apparent viscosity. Newtonian fluids are then characterized by different apparent viscosities at each shear velocity \cite{cl_4,cl_6}. 

Non-Newtonian fluids are mainly classified as: independent of the time and, dependent of the time. In fluids independent of time the viscosity at any shear stress does not vary with the time, while in dependents ones it does. Among the time independent of the time fluids are the pseudoplastics, dilators and Bingham fluids. The time dependent will not be studied in this article therefore it will not be deepened in them \cite{cl_1}.  

The behavior of viscosity for pseudoplastics and dilators fluids can be modeled mathematically using the the Power Law (Equation (\ref{power's_law})). On the other hand, if the exponent n of the equation is smaller than the unit ($n < 1$), it is called a pseudoplastic fluid, in the case where this exponent is larger than the unit ($n > 1$), it is called a dilating fluid \cite{libro_2}. The law of potency can also be used for Newtonian fluids, we have in this case the constant k is the viscosity and the value of $n = 1$ \cite{ref_libro}.


\section{Experimental set-up}

The set-up used for the experimental tests can be seen in Figure \ref{fig_2}. This is made up of a graduated cylinder of $500$ $ml$, a iron disk, a solid bronze sphere, and, two infrared sensors. The graduated cylinder has a mass of $458$ $g$, volume of $500$ $ml$ and radius of $23.35 \times 10^{-3}$ $m$. In order to prevent that the bronze sphere does not break the graduated cylinder in the fall, the iron disk is placed inside it. This disk has a mass of $68.5$ $g$ and a volume of $8.877 \times 10^{-6}$ $m^3$.  Once the iron disk is inside the graduated cylinder, this is filled with the fluid under study. 

As newtonian fluids were used: glycerine and sunflower oil; as non-newtonian fluids were used: yogurt, ketchup, cornstarch and corn flour. The sensors were located at a height $L$ of $0.017$ $m$. Once the fluid is inside, at room temperature, the mass sphere is launched. The sphere has a mass of $74.5$ $g$, volume of $8.877 \times 10^{-6}$ $m^3$, radius of $12.66 \times 10^{-3} m$ and a density of $2635.29$ $kg/m^{3}$.

The velocity of the sphere will be calculated as the distance L routed by the sphere, divided by the time that it takes to travel that distance (Equation (\ref{eq_vel})). The sensor will detect the passage of the sphere and record the time $\Delta t$. This time data collection was performed using the CASSY LAB-LD Didactic system software (as shown in Figure \ref{fig_2}). For each fluid, the sphere was launched 10 times. For each launch, the time taken by the sphere to travel the distance $L$ was measured by the detectors. The time of all releases was averaged. With the average time the velocity was calculated (See Equation (\ref{eq_vel})).

\begin{equation}
    v=\frac{L}{\Delta t}
    \label{eq_vel}
\end{equation}

\begin{Figure}
 \centering
 \includegraphics[width=0.8\linewidth]{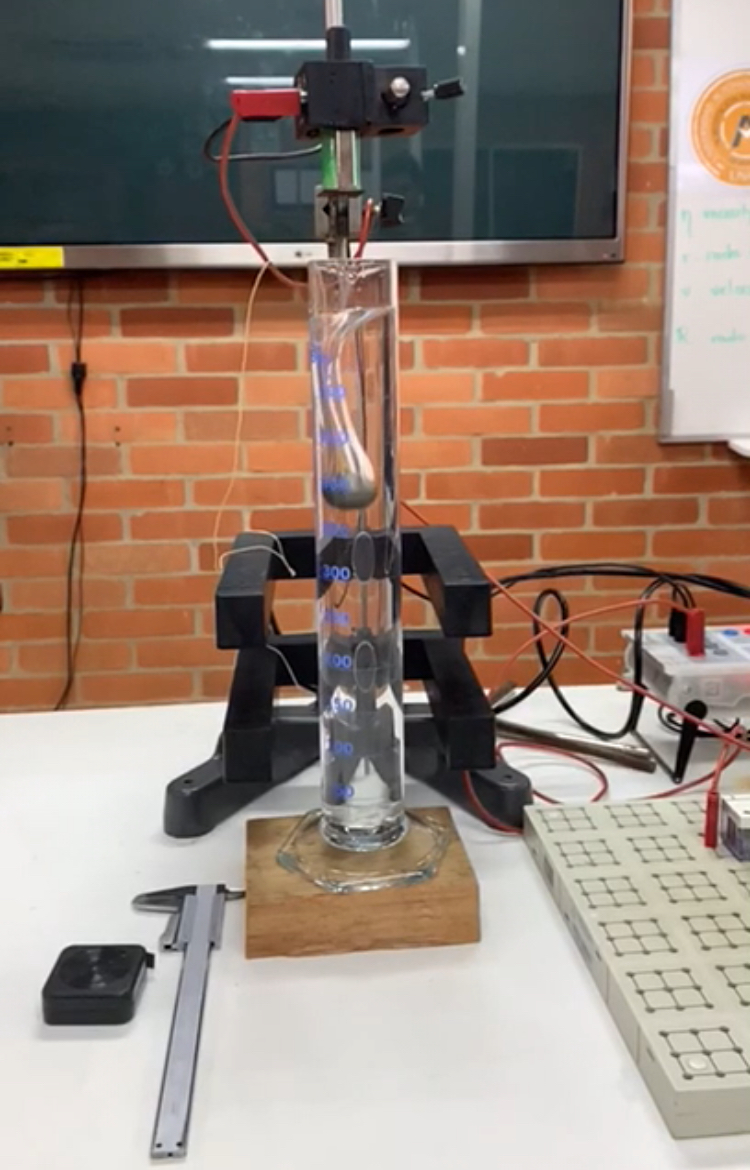}
 \captionof{figure}{Viscometer experimental set-up.}
    \label{fig_2}
\end{Figure}

The experimental viscosity, for each fluid was calculated as shown in Equation (\ref{eq_prin}), where $\rho _s$  is the sphere density, $\rho _f$ is the fluid density, $\Delta t$ the average time, $L$ the distance travelled, $r$ the sphere radius, $R$ the graduated cylinder radius and $g$ is the local gravity. Where the density of the mass of the iron disk and the density of the fluid were calculated from the measurement of its mass and volume. 

\begin{equation}
\eta = \left[ \frac{2}{9} (\rho _s - \rho _f) \left( \frac{\Delta t}{L} \right) r^2 \right] \cdot \left[ \frac{g}{1 + 2.4 \left( \frac{r}{R} \right)} \right]
    \label{eq_prin}
\end{equation}

In the case of non-Newtonian fluids, these were heated using a water bath as an external heat source at the bottom of the cylinder. In this case, data collection was performed as the fluid temperature increased. For these fluids type, the fall time $\Delta t$ of the bronze sphere as a function of temperature was measured, increasing it by $1$ to $2$ $^0$C on average from the ambient temperature ($25$ $^0$C). The number of time data, which were recorded in this case, based on a certain value of a temperature increase described above, from room temperature was $10$ times.

\section{Results}

To carry out the analysis of the results obtained in this research, we wanted to start by analyzing the non-Newtonian fluids that we have worked on in this research, which were:

\begin{enumerate}
    \item Yogurt.
    \vspace{-0.2cm}
    \item corn flour. 
     \vspace{-0.2cm}
    \item Ketchup.
     \vspace{-0.2cm}
    \item Cornstarch.
\end{enumerate}

On the other hand, we also work with two Newtonian fluids, which are:

\begin{enumerate}
    \item Sunflower oil.
     \vspace{-0.2cm}
    \item Glycerine.
\end{enumerate}

Taking into account this order of presentation of our results, we present below the results obtained for non-Newtonian yogurt fluid.

\subsection{Yogurt results analysis}

To begin with the analysis of the results obtained for this fluid, we have taken The average times obtained of the travel of the sphere in the distance $L$; at different temperatures are presented in the Table \ref{table_1}. With these times, the data of the fluid density, density and dimensions of the sphere were calculated, by replacing the viscosity in Equation (\ref{eq_prin}) (See Table \ref{table_1}). It was found that when the temperature increases, the time decreases, which represents a lower shear stress at a higher temperature, that is, a lower viscosity of the fluid. This variation in viscosity with temperature proves that ketchup sauce is a non-Newtonian fluid.

\begin{Table}
  \captionof{table}{Experimental data on the viscosity of Yogurt for different temperatures.}
   \begin{tabular}{ccc}
    \hline
   \textbf{T {[}K{]}} & \textbf{Average time {[}s{]}} & \textbf{Viscosity {[}Pa$\cdot$s{]}} \\ \hline \hline
293.05                  & 0.0694                         & 1.1317                      \\ 
295.05                  & 0.0683                        & 1.1131                      \\ 
298.35                  & 0.0666                        & 1.0855                      \\ 
304.15                  & 0.0653                        & 1.0651                      \\ 
311.75                  & 0.0648                        & 1.0561                       \\ 
315.15                  & 0.0635                        & 1.0349                       \\ \hline
   \end{tabular}
  \label{table_1}
\end{Table}

The results of the yogurt viscosity at different temperatures are presented in the Figure \ref{fig_4}a. Equation 10 was used to calculated the viscosity at each temperature, in this case the density of the yogurt used was 1052.6894 $kg/m^3$ \cite{ref_fig_1}. Because yogurt is a non-Newtonian fluid, a exponential adjustment was fitted, obtaining a good approximation with an $R^2$ of $0.929$. The model obtained corresponds to the Arrhenius's equation  (See Equation (\ref{eq_11})), In which a dependence of the viscosity of the fluid with temperature can be observed, in which a notable decrease in viscosity can be evidenced as the temperature increases, in view of the shape of this figure  (See Figure \ref{fig_4}a), indicating that the fluid is pseudoplastic.

\begin{equation}
   \eta(T) = \eta_0 \cdot e^{\frac{E}{RT}},
    \label{eq_11}
\end{equation}

where $\eta$ is the viscosity of fluid, $\eta_0$ is a pre-exponential factor, $E$ is the activation energy, $R$ is the ideal gases universal constant and $T$ is the absolute temperature of fluid.

\begin{Figure}
 \centering
 \includegraphics[width=1.0\linewidth]{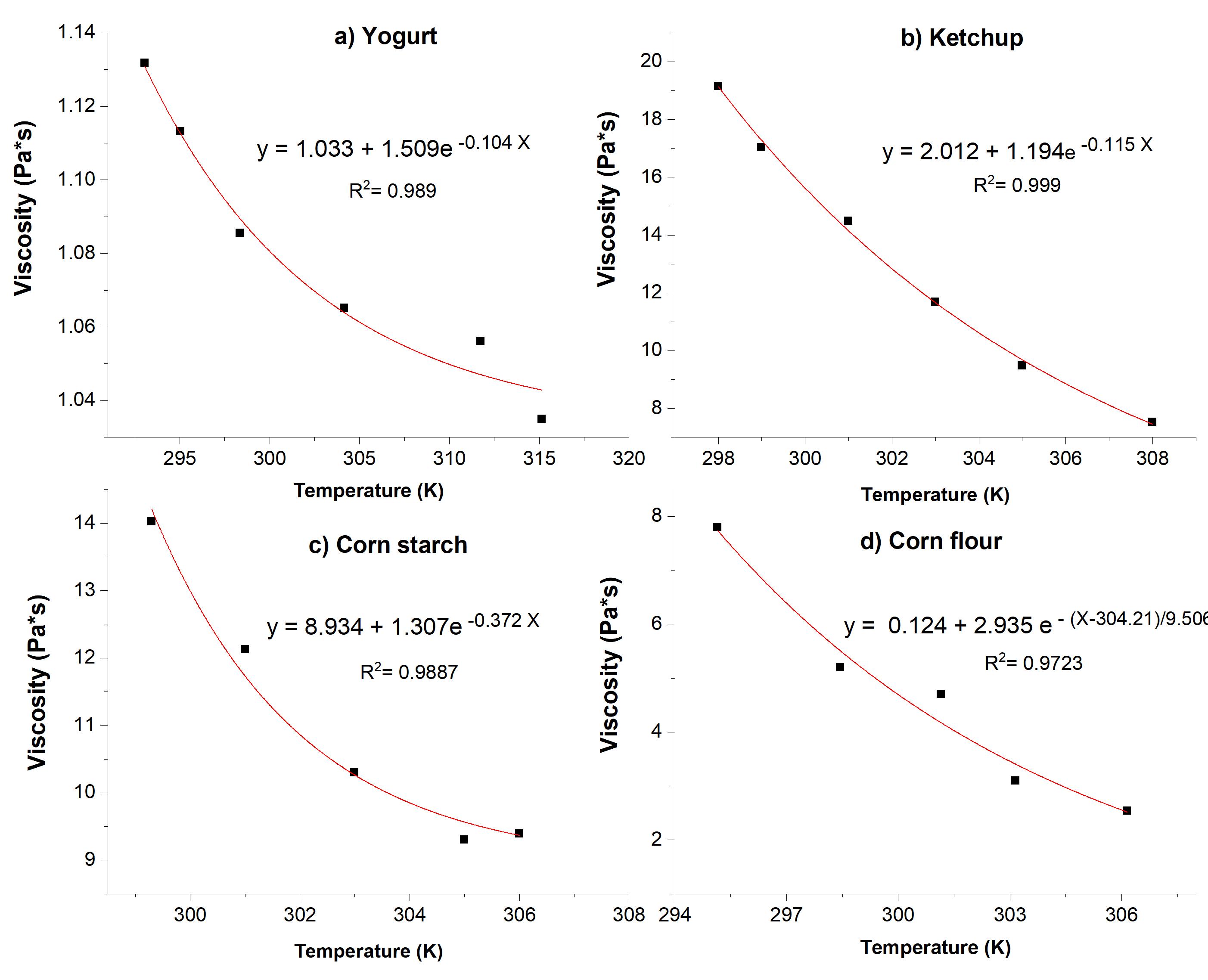}
 \captionof{figure}{Viscosity as a function of temperature for non-Newtonian fluids.a.) Yogurt. b.) Corn flour. c.) Ketchup. d.) Cornstarch. }
    \label{fig_4}
\end{Figure}

Now, we can also perform a linear adjustment of the Arrhenius's equation (See Equation (\ref{eq_11})), applying the method of least squares, where Equation (\ref{eq_11}), we can linearize it as follows:

\begin{align}
       \ln [\eta(T)] &= \ln \left[ \eta_0 \cdot e^{\frac{E}{RT}} \right], \\
       \ln [\eta(T)] &= \ln [\eta_0] + \ln \left[ e^{\frac{E}{RT}} \right]  \\
       \ln [\eta(T)] &= \ln [\eta_0] + \left[ \frac{E}{R} \right] \cdot \dfrac{1}{T}, \label{eq_14} \\
       y &= b + m \cdot x. \label{eq_15}
\end{align}

In this way, comparing Equation (\ref{eq_14}) with Equation (\ref{eq_15}), it can be seen that making the graph of $\ln \eta$ as a function of $[1/T]$, where moreover making a linear adjustment (See Figure \ref{fig_5}), the value of the activation energy $E$ of the fluid can be obtained experimentally

\begin{Figure}
 \centering
 \includegraphics[width=1.0\linewidth]{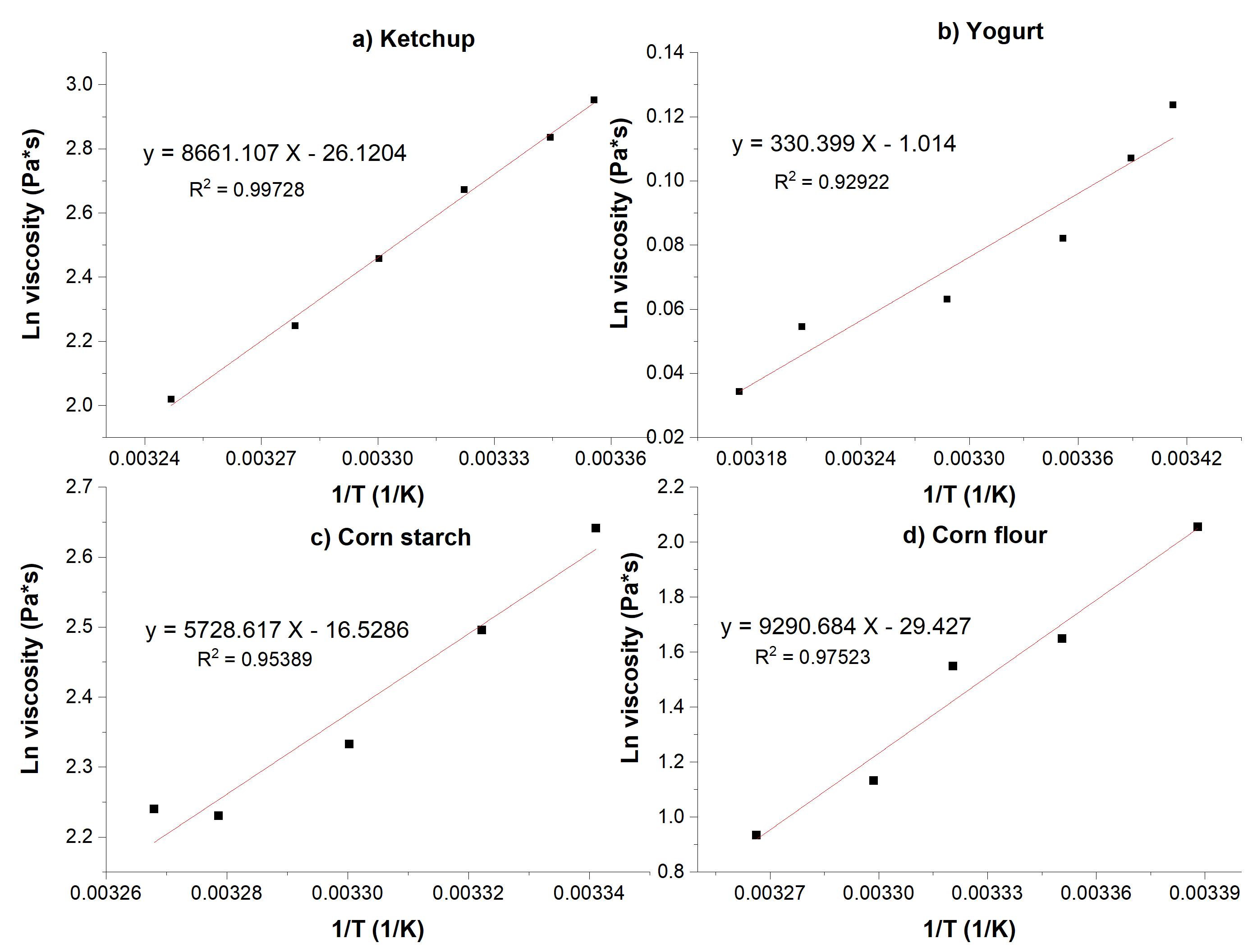}
 \captionof{figure}{Linear fitting of viscosity  as a function of temperature for non-Newtonian fluids. a.) Yogurt. b.) Corn flour. c.) Ketchup. d.) Cornstarch. }
    \label{fig_5}
\end{Figure}

From the linearization used in Equation (\ref{eq_14}) to power the linear fit, shown in Figure \ref{fig_5}b, the value of the activation energy $E$ can be calculated, in the case of yogurt, as can be seen in the equations from (\ref{eq_16}) to (\ref{eq_17}).

\begin{align}
     \frac{E}{R} &= m, \label{eq_16} \\
    E &= m \cdot R. \label{eq_17} 
\end{align}
   
Where m is the slope of Figure \ref{fig_5}b, which corresponds to a numerical value of $m = 330.399$ $K$, with this value; and using the value of the universal constant of ideal gases given by the unit literature in the I.S system, we have $R = 8.314$ $J/(K\cdot mol)$, we obtain the result of the activation energy $E$, shown in equations (\ref{eq_18}-\ref{eq_19}), 

\begin{align}
    E &= 330.399 \cdot 8.314, \label{eq_18} \\
    E &= 2.747 \quad kJ/mol. \label{eq_19}    
\end{align}

Continuing with the presentation of the results for Yogurt, we have constructed Table \ref{table_2}, which shows the calculation of the Reynolds number using Equation (\ref{reinolds}); for different temperature values.

\begin{Table}
  \captionof{table}{Experimental data from the calculation of the Reynolds number for yogurt as a function of the falling velocity of the sphere.}
   \begin{tabular}{ccc}
    \hline
   \textbf{Viscosity {[}Pa$\cdot$s{]}} & \textbf{Velocity $v$ {[}m/s{]}} & \textbf{Re} \\ \hline \hline
1.1317                  & 0.8645                         & 10.8559                      \\ 
1.1131                 & 0.8784                        & 11.2148                      \\ 
1.0855                  & 0.9009                        & 11.7945                      \\ 
1.0651                 & 0.9188                        & 12.2592                      \\ 
1.0561                  & 0.9259                        & 12.4592                       \\ 
1.0349                  & 0.9448                        & 12.9740                       \\ \hline
   \end{tabular}
  \label{table_2}
\end{Table}

In this way, we can verify that the obtained Reynolds number (See Table \ref{table_2}), is between ($0.8645 <Re<0.9448$), for which it can be assumed that the flow remains stationary and behaves as if it were formed by sheets thin ($Re <2300$), which interact only according to the existing tangential stresses. So this flow is called laminar flow.

\subsection{Corn flour results and analysis}

The experimental density that was obtained for the corn flour was $602.9$ $kg/m^3$. The experimental data of the average time and viscosity for different temperatures are shown in Table \ref{table_3}.

\begin{Table}
  \captionof{table}{Experimental data on the viscosity of corn flour for different temperatures.}
   \begin{tabular}{ccc}
    \hline
   \textbf{T {[}K{]}} & \textbf{Average time {[}s{]}} & \textbf{Viscosity {[}Pa$\cdot$s{]}} \\ \hline \hline
295.15                  & 0.7405                         & 7.7974                      \\ 
298.45                  & 0.4934                        & 5.1954                      \\ 
301.15                  & 0.4466                        & 4.7026                      \\ 
303.15                  & 0.2944                        & 3.1000                     \\ 
306.15                  & 0.2413                        & 2.5414                       \\           \hline
   \end{tabular}
  \label{table_3}
\end{Table}

In Figure \ref{fig_4}d, the graph of viscosity vs. temperature was made, where we can see how viscosity varies with temperature; with a function, whose mathematical expression is an exponential with negative index. Obtaining a mathematical expression, to model the behavior of viscosity as a function of temperature, for the case of corn flour (non-Newtonian fluid). This is shown in Figure \ref{fig_4}d.

In this way we can see that with a Pearson regression factor $R^2 = 0.9723$ approximately, we can say that this model fits the experimental data obtained quite well (See Figure \ref{fig_4}d). This behavior is to be expected according to the literature. Where it can be noted that as the temperature value increases, the viscosity value decreasing, reflected in an increase in the falling velocity of the sphere inside the fluid (See Figure \ref{fig_4}d). In Figure \ref{fig_4}d, you can also see a recent behavior; on the curve of viscosity as a function of temperature, which indicates that the fluid is pseudoplastic modeled by Equation \ref{eq_11}.

On the other hand, fitting the data to a linear model (See Figure \ref{fig_5}d), the activation energy $E$ of corn flour can be calculated, using the linearization of Equation (\ref{eq_14}) and Equation (\ref{eq_15}), with these Two equations, the result shown in Equation (\ref{eq_21}) of the energy $E$ is obtained for corn flour.

\begin{align}
    E &= 9290.684 \cdot 8.314, \label{eq_20} \\
    E &= 77.242 \quad kJ/mol. \label{eq_21}    
\end{align}

Continuing with the presentation of the results, obtained for corn flour, we can calculate the value of the Reynolds number (Re), given for each of the different temperatures, at which this experiment was carried out (See Equation (\ref{reinolds})) . In Table \ref{table_4}, we show the value of the fluid viscosity (corn flour), as a function of the falling velocity $v$, for each of the tests carried out. The results of these experimental calculations are presented in Table \ref{table_4}.

\begin{Table}
 \captionof{table}{Experimental data from the calculation of the Reynolds number for corn fluor as a function of the falling velocity of the sphere.}
   \begin{tabular}{ccc}
    \hline
   \textbf{Viscosity {[}Pa$\cdot$s{]}} & \textbf{Velocity $v$ {[}m/s{]}} & \textbf{Re} \\ \hline \hline
7.7974                 & 0.1350                         & 0.1409                      \\ 
5.1954                 & 0.2026                        & 0.3175                      \\ 
4.7026                 & 0.2239                        & 0.3875                      \\ 
3.1000                 & 0.3396                        & 0.8918                      \\ 
2.5414                  & 0.4143                        & 1.3269                      \\ \hline
   \end{tabular}
  \label{table_4}
\end{Table}

In this way (See Table \ref{table_4}), we can verify that the Reynolds number obtained is within the following range ($0.1409<Re<1.3269$), for which it can be assumed that the flow remains stationary and behaves as if it were formed by thin sheets, which interact only depending on the existing tangential stresses, so this flow is called laminar flow in which the Reynolds number value is $Re<2300$.

\subsection{Ketchup results analysis}

The average times obtained of the travel of the sphere in the distance $L$ at different temperatures are presented in Table \ref{table_5}. With these times, the data of the fluid density, dimensions and density of the sphere were calculated by replacing the viscosity in Equation (\ref{eq_prin}) (See Table \ref{table_5}). It was found that when the temperature increases, the time decreases, which represents a lower shear stress at a higher temperature, that is, a lower viscosity of the fluid. This variation in viscosity with temperature proves that ketchup sauce is a non-Newtonian fluid.

\begin{Table}
  \captionof{table}{Experimental data on the viscosity of ketchup for different temperatures.}
   \begin{tabular}{ccc}
    \hline
   \textbf{T {[}K{]}} & \textbf{Average time {[}s{]}} & \textbf{Viscosity {[}Pa$\cdot$s{]}} \\ \hline \hline
298                  & 1.677                         & 19.14                      \\ 
299                  & 1.492                        & 17.04                      \\ 
301                  & 1.268                        & 14.48                      \\ 
303                  & 1.023                        & 11.68                     \\ 
305                  & 0.786                        & 9.47                       \\  
308                  & 0.659                        & 7.53                       \\           \hline
   \end{tabular}
  \label{table_5}
\end{Table}

The results of the ketchup viscosity at different temperatures are presented in the Figure \ref{fig_4}b. Equation (\ref{eq_prin}) was used to calculated the viscosity at each temperature, in this case the density of the ketchup used was $1235$ $kg/m^3$ \cite{ref_fig_1}. Because ketchup is a non-Newtonian fluid, a power adjustment was fitted, obtaining a good approximation with an $R^2$ of $0.999$ (See Figure \ref{fig_4}b).  The model obtained corresponds to the exponential model (See Equation (\ref{eq_11})), In Figure \ref{fig_4}b, it is shown that the curve for the behavior of viscosity as a function of temperature is a decreasing exponential behavior as shown in Equation (\ref{eq_11}), in which it is evident that ketchup corresponds to a pseudoplastic fluid.

On the other hand, fitting the data to a linear model (See Figure \ref{fig_5}a), the activation energy $E$ of ketchup can be calculated, using the linearization of Equation (\ref{eq_14}) and Equation (\ref{eq_15}), with these Two equations, the result shown in Equation (\ref{eq_23}) of the energy $E$ is obtained for ketchup.

\begin{align}
    E &= 8861.107 \cdot 8.314, \label{eq_22} \\
    E &= 73.671 \quad kJ/mol. \label{eq_23}    
\end{align}

On the other hand, the Reynolds number can be calculated as a function of the viscosity using the expression of the Equation (\ref{reinolds}). In the Table \ref{table_6}, shows the Reynolds number for each temperature and each velocity reached for the sphere in the ketchup fluid.

\begin{Table}
  \captionof{table}{Experimental data from the calculation of the Reynolds number for ketchup as a function of the falling velocity of the sphere.}
   \begin{tabular}{ccc}
    \hline
   \textbf{Viscosity {[}Pa$\cdot$s{]}} & \textbf{Velocity $v$ {[}m/s{]}} & \textbf{Re} \\ \hline \hline
19.14                 & 0.0495                         & 0.1491                      \\ 
17.04                 & 0.0556                        & 0.1676                      \\ 
14.48                 & 0.0654                        & 0.1972                      \\ 
11.68                 & 0.0811                        & 0.2443                      \\ 
9.47                  & 0.1057                        & 0.3014 
                \\
7.53                 & 0.1259                        & 0.3792                  \\ \hline
   \end{tabular}
  \label{table_6}
\end{Table}

\subsection{Cornstarch results analysis}

The average times obtained of the travel of the sphere in the distance $L$ at different temperatures are presented in Table \ref{table_7}. With these times, the data of the fluid density, density and dimensions of the sphere were calculated by replacing the viscosity in Equation (\ref{eq_prin}) (See Table \ref{table_7}). It was found that when the temperature increases, the time decreases, which represents a lower shear stress at a higher temperature, that is, a lower viscosity of the fluid. This variation in viscosity with temperature proves that cornstarch sauce is a non-Newtonian fluid.

\begin{Table}
  \captionof{table}{Experimental data on the viscosity of cornstarch for different temperatures.}
   \begin{tabular}{ccc}
    \hline
   \textbf{T {[}K{]}} & \textbf{Average time {[}s{]}} & \textbf{Viscosity {[}Pa$\cdot$s{]}} \\ \hline \hline
299                  & 1.1784                         & 14.025                      \\ 
300                  & 1.1784                        & 14.025                      \\ 
303                  & 1.019                        & 12.127                      \\ 
305                  & 0.8284                        & 9.855                     \\ 
305                  & 0.8277                        & 9.850                       \\  
306                  & 0.8026                        & 9.552                       \\           \hline
   \end{tabular}
  \label{table_7}
\end{Table}

The results of the ketchup viscosity at different temperatures are presented in the Figure \ref{fig_4}c. Equation (\ref{eq_prin}) was used to calculated the viscosity at each temperature, in this case the density of the cornstarch used was $1207,31$ $kg/m^3$ \cite{ref_fig_1}. Because cornstarch is a non-Newtonian fluid, a power adjustment was fitted, obtaining a good approximation with an $R^2$ of $0.9887$ (See Figure \ref{fig_4}c).  The model obtained corresponds to the exponential model (See Equation (\ref{eq_11})), In Figure \ref{fig_4}c, it is shown that the curve for the behavior of viscosity as a function of temperature is a decreasing exponential behavior as shown in Equation (\ref{eq_11}), in which it is evident that cornstarch corresponds to a fluid that is pseudoplastic.

On the other hand, fitting the data to a linear model (See Figure \ref{fig_5}c), the activation energy $E$ of cornstarch can be calculated, using the linearization of Equation (\ref{eq_14}) and Equation (\ref{eq_15}), with these two equations, the result shown in Equation (\ref{eq_25}) of the energy $E$ is obtained for cornstarch.

\begin{align}
    E &= 5728.617 \cdot 8.314, \label{eq_24} \\
    E &= 47.627 \quad kJ/mol. \label{eq_25}    
\end{align}

On the other hand, the Reynolds number can be calculated as a function of the viscosity using the expression of the Equation (\ref{reinolds}). Table \ref{table_8}, shows the Reynolds number for each temperature and each velocity reached for the sphere in the cornstarch fluid.

\begin{Table}
  \captionof{table}{Experimental data from the calculation of the Reynolds number for cornstarch as a function of the falling velocity of the sphere.}
   \begin{tabular}{ccc}
    \hline
   \textbf{Viscosity {[}Pa$\cdot$s{]}} & \textbf{Velocity $v$ {[}m/s{]}} & \textbf{Re} \\ \hline \hline
14.025                 & 0.0678                         & 0.2729                      \\ 
14.025                 & 0.0679                        & 0.2728                      \\ 
12.127                 & 0.0785                        & 0.3650                      \\ 
9.855                 & 0.0965                        & 0.5285                      \\ 
9.850                  & 0.0966                        & 0.5858 
                \\
9.552                 & 0.0996                        & 0.5984                  \\ \hline
   \end{tabular}
  \label{table_8}
\end{Table}

\subsection{Sunflower oil results and analysis}

In the case of sunflower oil, where it has a Newtonian fluid behavior, in which with increasing temperature, the viscosity value will remain constant. It is for this reason that we only take the fall time of the sphere, for a single temperature, which in our case was $297$ $K$. Using Equation (\ref{eq_prin}), for an experimentally measured density value equal to $907.42$ $kg/m^3$, a viscosity value was obtained for sunflower oil, equal to $41.22$ $\times 10^{-3}$ $Pa\cdot s$ \cite{al_1,al_2}. Carrying out the respective comparison, with the theoretical value reported by the literature, we found a percentage error of $5.57$ $\%$, in which a great closeness with the theoretical value could be evidenced; which is evidence that an experimental procedure could be successfully carried out, with the purpose of indirectly measuring the viscosity of sunflower oil; this being a Newtonian fluid.

\subsection{Glycerine results and analysis}

For the Newtonian fluid glycerine, $490.928$ $mL$ of glycerine were taken in the cylinder, this volume was weighed obtaining a mass of $1.115$ $kg$. The density was calculated by dividing the mass of glycerine in the volume, obtaining an experimental density of $1177.374$ $kg/m^3$.
The distance traveled by the sphere was $0.1065$ $m$ and the average time was $0.1578$ $s$. With these values the velocity was calculated obtaining $0.6748$ $m/s$. These values were replaced in Equation (\ref{eq_prin}) and the viscosity value was calculated, obtaining an experimental value of $1.4183$ $Pa\cdot s$. The reported viscosity value for commercial glycerine at a temperature of $298$ $K$ is $1.412$ $Pa \cdot s$ \cite{cl_7,cl_8}, which indicates that the experimental value obtained in this study is very close to the reported theoretical value.

\section{Acknowledgments}

The authors: Alex Estupi\~n\'an and Claudia Tavera, would like to express their thanks, especially to the Universidad de Investigaci\'on y Desarrollo \textit{UDI}, for all the human, material and financial support to carry out this research work. Also the authors: Ra\'ul Rojas and Juan Quijano, they want to thank the Universidad Aut\'onoma de Bucaramanga \textit{UNAB}, for providing us with technical and financial support.

\section{Conclusions}

In this work was possible to implement an experimental protocol to perform the indirect measurement of the viscosity of Newtonian and non-Newtonian fluids. The results obtained in this research are shown with high precision and accuracy, and it is possible to catalog the behaviour of the different fluids.

It was demonstrated that our prototype can accurately measure the viscosity of both Newtonian and non-Newtonian fluids, in addition to working properly at different temperature conditions. 

Additionally, due to the good adjustments presented in the graphs of $1/T$ Vs Ln of viscosity, it was possible to calculate the transition energy of the molecules, which are in accordance with the values presented in the literature. Likewise, experimental values of the Reynolds number could be obtained, with which it is possible to predict the behavior of the fluid. In this case, the fluids were in steady flow, which could be verified with the $Re$ values obtained

\end{multicols}

\end{document}